\documentclass[nofootinbib,prd]{revtex4}%
\usepackage{amsmath}
\usepackage{amsfonts}
\usepackage{amssymb}
\usepackage{graphicx}%
\usepackage{amsmath,amssymb}
\usepackage{mathrsfs}
\usepackage{graphicx}
\usepackage{color}
\usepackage{subfigure}
\usepackage{fancyhdr}
\usepackage{multirow}
\usepackage{float}
\usepackage{epsfig}
\usepackage{amsfonts}
\usepackage{bm}

\begin{document}
\title{Traversable Wormholes supported by Holographic Dark Energy with a modified Equation of State}
\author{Remo Garattini}
\email{remo.garattini@unibg.it}
\affiliation{Universit\`a degli Studi di Bergamo, Dipartimento di Ingegneria e Scienze
Applicate, Viale Marconi 5, 24044 Dalmine (Bergamo) Italy and I.N.F.N.-
sezione di Milano, Milan, Italy.}
\author{Phongpichit Channuie}
\email{phongpichit.ch@mail.wu.ac.th}
\affiliation{College of Graduate Studies, Walailak University, Thasala, \\Nakhon Si Thammarat, 80160, Thailand\\School of Science, Walailak University, Thasala, \\Nakhon Si Thammarat, 80160, Thailand}

\begin{abstract}
Inspired by holographic dark energy models, we consider different energy density profiles as possible sources needed to have traversable wormholes solutions. Since such energy densities are all positive,
we are forced to introduce an equation of state of the form $p_{r}%
(r)=\omega_{r}\left(  r\right)  \rho(r)$. We will find that Zero Tidal Forces
can be imposed at the price of having the function $\omega_{r}\left(
r\right)  $ divergent for $r\rightarrow\infty$. To overcome this inconvenient, we abandon the request of having Zero Tidal Forces, by introducing appropriate modifications on the function $\omega_{r}\left(  r\right)  $ in such a way to obtain a finite result everywhere. We will find that such modifications will leave the behavior of the Equation of State close to the throat invariant. Moreover, despite of the initial assumption, we will find that every dark energy profile will be moved into the phantom region. Among the different energy density proposals, only one profile will not require a modification of the original $\omega
_{r}\left(r\right)  $ to have Zero Tidal Forces. Such an energy density profile will be
consistent with the appearance of a Global Monopole.

\end{abstract}
\maketitle

\section{Introduction}

The null energy condition (NEC) establishes that for any null vector
$T_{\mu\nu}k^{\mu}k^{\nu}\geq0$. $T_{\mu\nu}$ represents the Stress-Energy
Tensor (SET). In terms of the energy density $\rho$ and pressures $p_{i}$, the
NEC can be written as%
\begin{equation}
\rho+p_{i}\geq0\qquad i=1,2,3.
\end{equation}
A Traversable Wormhole (TW) is a solution of the Einstein Field Equations
(EFE) with the property of connecting two distant regions by means of a tunnel
violating the NEC. In particular, if we indicate with $p_{r}$ the radial
pressure, the following inequality%
\begin{equation}
\rho+p_{r}\leq0
\end{equation}
must hold. The first steps moved towards a TW analysis have been described by
Ludwig Flamm\cite{Flamm:1916} and subsequently by Einstein and Rosen
\cite{Einstein:1935tc}, where the Schwarzschild solution has been cast into a
form such that to describe a bridge-like solution: the so-called
Einstein-Rosen (ER) bridge. We have to wait for Morris and Thorne \cite{MT,Morris:1988tu} to have the modern formulation for a
TW \cite{Visser1995}. It is interesting to note that the Schwarzschild solution
represents a wormhole which is not traversable. Note also that the NEC
violation is related to the existence of \textquotedblleft
exotic\textquotedblright\ matter. Hence, the semiclassical theory (or perhaps
a possible quantum theory) of gravity may be a good tool to tackle the
underlying descriptions of a TW. The Casimir energy is the most likely
candidate of matter forms that can be used to stabilize a TW. It is one type
of the vacuum energy which has been confirmed by the experiment. There exists
an extension which underlies a so-called Generalized Uncertainty Principle
(GUP) \cite{Jusufi:2020rpw} including the electric charge
\cite{Samart:2021tvl,Garattini:2023qyo}. Moreover, related investigations on the Casimir
effect have been noticed in the literature
\cite{EPJC1,Garattini:2020kqb,Garattini:2021kca,Santos:2021jjs}. Casimir wormholes have been also studied in the context of modified theories of gravity, see e.g., \cite{Hassan:2022hcb,Sokoliuk:2022jcq,Oliveira:2021ypz}.

Models containing a component with an arbitrary equation of state
$\omega=p/\rho$ is known as the dark energy. Values of $\omega<-1/3$ are
required for cosmic acceleration. For $\omega<-1$, one enters into the realm of
phantom energy. Although there are a lot of models and theories having been
proposed to understand the nature of dark energy, they are still far to be satisfying. However, among numerous dynamical dark energy
models, a promising approach comes from the \textquotedblleft holographic principle\textquotedblright \cite{tHooft:1993dmi,thooft2009,Susskind1995,Susskind2004,Maldacena1999,Cohen1999}, see also a review \cite{Bousso2002}. It refers to the duality between theories of
the bulk and its boundary. The
application of the holographic principle to study the nature of dark energy
has been widely regarded as an attractive approach. It is noticed that the
holographic principle is inspired by the investigations of quantum properties
of black holes and shed some light on the cosmological problem and the dark
energy problem. In the holographic principle (HP), the conjecture proposed in Ref.\cite{Cohen1999} establishes a connection between energy density and length scale. In principle, one can identify the dark energy density as the vacuum energy density of the underlying effective field theory and proposes a dynamical expression as $\rho_{\Lambda}\propto S/L^{4}$ with $S$ being the entropy of the cosmological
horizon and the infrared cut-off, $L$, relevant to the dark energy is the size of the
event horizon. Accounting HP, the observed density of dark energy, $\rho_{\Lambda}$, might be possibly explained, see for example \cite{SayahianJahromi:2018irq}. In Ref.\cite{Li:2004rb}, the author proposed the Holographic Dark Energy (HDE) model as
$\rho_{\rm Holo}=3c^{2}M_{p}^{2}L^{-2}$ with $c$ being a numerical factor. In case of black hole (BH) physics, it is proportional to the BH
surface area,
\cite{Bekenstein1973,Bekenstein1974,Hawking1974,Hawking1975,Hawking1976,Gibbons1977,Unruh1976,Bardeen1973}. One can think that the apparent horizon can be promoted to a
radial coordinate. In this case, we can think of an inspired energy density
and hence particular kind of an equation of state (EoS) to be used as a
potential source of a traversable wormholes (TWs). Units in
which $\hbar=c=k=1$ are used throughout the paper and will be reintroduced
whenever it is necessary.

\section{General Setup}

\label{ch2}
We consider a static and spherically symmetric Morris-Thorne traversable
wormhole in the Schwarzschild coordinates given by \cite{MT}
\begin{equation}
ds^{2}=-e^{2\Phi(r)}dt^{2}+\frac{dr^{2}}{1-\frac{b(r)}{r}}+r^{2}\left(
d\theta^{2}+\sin^{2}\theta d\phi^{2}\right)  ,\label{ds}%
\end{equation}
in which $\Phi(r)$ and $b(r)$ are the redshift and shape functions,
respectively. In the wormhole geometry, the redshift function $\Phi(r)$ should
be finite in order to avoid the formation of an event horizon. Moreover, the
shape function $b(r)$ determines the wormhole geometry, with the following
condition $b(r_{0})=r_{0}$, in which $r_{0}$ is the radius of the wormhole
throat. Consequently, the shape function must satisfy the flaring-out
condition \cite{MT}:
\begin{equation}
\frac{b(r)-rb^{\prime}(r)}{b^{2}(r)}>0,
\end{equation}
in which $b^{\prime}(r_{0})<1$ must hold at the throat of the wormhole. With
the help of the line element $\left(  \ref{ds}\right)  $, we obtain the
following set of equations resulting from the energy-momentum components to
yield $\left(  \kappa=8\pi G\right)  $%
\begin{equation}
\frac{b^{\prime}(r)}{r^{2}}=\kappa\rho(r),\label{rho}%
\end{equation}%
\begin{equation}
\left[  2\left(  1-\frac{b(r)}{r}\right)  \frac{\Phi^{\prime}(r)}{r}%
-\frac{b(r)}{r^{3}}\right]  =\kappa p_{r}(r)\label{pr}%
\end{equation}
and%
\begin{align}
&  \Bigg\{\left(  1-\frac{b\left(  r\right)  }{r}\right)  \left[  \Phi
^{\prime\prime}(r)+\Phi^{\prime}(r)\left(  \Phi^{\prime}(r)+\frac{1}%
{r}\right)  \right]  \nonumber\\
&  -\frac{b^{\prime}\left(  r\right)  r-b\left(  r\right)  }{2r^{2}}\left(
\Phi^{\prime}(r)+\frac{1}{r}\right)  \Bigg\}=\kappa p_{t}(r),\label{pt}%
\end{align}
where $p_{t}=p_{\theta}=p_{\phi}$. We can complete the EFE with the expression
of the conservation of the Stress-Energy Tensor (SET) which can be written in
the same orthonormal reference frame%
\begin{equation}
p_{r}^{\prime}\left(  r\right)  =\frac{2}{r}\left(  p_{t}\left(  r\right)
-p_{r}\left(  r\right)  \right)  -\left(  \rho\left(  r\right)  +p_{r}\left(
r\right)  \right)  \Phi^{\prime}(r).\label{Tmn}%
\end{equation}
If Zero Tidal Forces (ZTF) are considered, we can impose the following
Equation of State (EoS)%
\begin{equation}
p_{r}(r)=\omega_{r}\left(  r\right)  \rho(r)\label{EoS1}%
\end{equation}
and write $\Phi(r)=0$. This is equivalent to%
\begin{equation}
\omega_{r}\left(  r\right)  =-\frac{b(r)}{rb^{\prime}(r)}.\label{or(r)}%
\end{equation}
Solving the previous equation with respect to $b(r)$, one finds the well know
profile%
\begin{equation}
b(r)=r_{0}\,\exp\left[  -\int_{r_{0}}^{r}\,\frac{d\bar{r}}{\omega_{r}\left(
\bar{r}\right)  \bar{r}}\right]  \,\label{form}%
\end{equation}
which must be consistent with the result obtained by solving Eq.$\left(
\ref{rho}\right)  $. It is straightforward to see that some energy density
profiles produce a divergent $\omega_{r}\left(  r\right)  $ to keep the
validity of ZTF. For instance, Casimir wormholes\cite{EPJC1} and
Yukawa-Casimir wormholes\cite{Garattini:2021kca} fall in this case, while the
Ellis-Bronnikov TW does not. Indeed, the Ellis-Bronnikov is described by the
following shape function%
\begin{equation}
b(r)=\frac{r_{0}^{2}}{r}%
\end{equation}
and $\omega_{r}\left(  r\right)  =1$, while the Casimir wormhole has a shape
function%
\begin{equation}
b(r)=\frac{2}{3}r_{0}+\frac{r_{0}^{2}}{3r}.
\end{equation}
and the associated $\omega_{r}\left(  r\right)  $ is such that%
\begin{equation}
\omega_{r}\left(  r\right)  =\left(  \frac{2r}{r_{0}}+1\right)  \underset
{r\rightarrow\infty}{\rightarrow}\infty.
\end{equation}
A similar behavior holds also for Yukawa-Casimir wormholes. In this paper we
would like to apply a modification of the original inhomogeneous EoS in such a
way to convert divergent $\omega_{r}\left(  r\right)  $ for large $r$ into a
convergent function. In particular, we are interested in HDE profiles which
will be described in the next section.

\section{Wormhole Geometries with holographic dark energy densities}

\label{ch3}We are going to apply the general setup presented in the previous
section to some specific energy density profiles inspired by HDE. They are:

\begin{enumerate}
\item Bekenstein-Hawking (BH) HDE with an energy density of the form%
\begin{equation}
\rho_{1}\left(  r\right)  =\frac{C\pi}{r^{2}}, \label{BH}%
\end{equation}

\item Moradpour et al.'s proposal with the following energy density profile%
\begin{equation}
\rho_{2}\left(  r\right)  =\frac{C}{4\pi r^{2}\left(  \pi\lambda
r^{2}+1\right)  }=\frac{C}{4\pi}\left(  \frac{1}{r^{2}}-\frac{\pi\lambda}%
{\pi\lambda r^{2}+1}\right)  , \label{M-rho}%
\end{equation}

\item Standard Renyi HDE with the following energy density form
\begin{equation}
\rho_{3}\left(  r\right)  =\frac{C}{\lambda r^{4}}\ln\left(  1+\pi\lambda
r^{2}\right)  , \label{R-rho}%
\end{equation}

\item Mixed Energy density%
\begin{equation}
\rho_{4}\left(  r\right)  =\frac{3C_{M}^{2}}{8\pi^{2}}\left[  \frac{\pi}%
{r^{2}}-\pi^{2}\lambda\ln\left(  1+\frac{1}{\pi\lambda r^{2}}\right)  \right]
.
\end{equation}

\end{enumerate}

As already mentioned in the previous section, it is likely that some profiles
can produce a divergent $\omega_{r}\left(  r\right)  $ when $r\rightarrow
\infty$. In such a case the choice $\left(  \ref{or(r)}\right)  $ must be
modified in an appropriate manner. For instance, if%
\begin{equation}
\omega_{r}\left(  r\right)  \sim r^{\alpha},\text{ for }r\rightarrow
\infty\text{ and }\alpha>0,
\end{equation}
then we can define%
\begin{equation}
\omega_{r}\left(  r\right)  =-\frac{r_{0}^{\alpha}b(r)}{r^{1+\alpha}b^{\prime
}(r)},\label{or(r)mod}%
\end{equation}
which is now convergent for $r\rightarrow\infty$. Of course, with such a
modification the behavior on the throat is unaffected and the ZTF cannot be
imposed. Rather we have to determine the form of the redshift function. To
this purpose, plugging Eq.$\left(  \ref{or(r)mod}\right)  $ into Eq.$\left(
\ref{pr}\right)  $, we obtain%
\begin{equation}
\left[  2\left(  1-\frac{b(r)}{r}\right)  \frac{\Phi^{\prime}(r)}{r}%
-\frac{b(r)}{r^{3}}\right]  =-\frac{r_{0}^{\alpha}b(r)}{r^{3+\alpha}},
\end{equation}
which can be rearranged to give%
\begin{equation}
\Phi^{\prime}(r)=\frac{b(r)\left(  r^{\alpha}-r_{0}^{\alpha}\right)
}{2r^{\alpha+1}\left(  r-b(r)\right)  }=\frac{b(r)\left(  r-r_{0}\right)
}{2r^{\alpha+1}\left(  r-b(r)\right)  }\sum_{i=0}^{\alpha-1}(r^{\alpha
-i-1}r_{0}^{i}).\label{Phi'(r)Mo}%
\end{equation}
Close to the throat, we can use the following approximation%
\begin{equation}
b\left(  r\right)  \simeq r_{0}+B\left(  r-r_{0}\right)  +O\left(  \left(
r-r_{0}\right)  ^{2}\right)  ,\label{b(r)t}%
\end{equation}
where $B=b^{\prime}(r_{0})$. Then Eq.$\left(  \ref{Phi'(r)Mo}\right)  $
becomes%
\begin{equation}
\Phi^{\prime}(r)\simeq\frac{r_{0}}{2r^{\alpha+1}\left(  1-B\right)  }%
\sum_{i=0}^{\alpha-1}(r^{\alpha-i-1}r_{0}^{i}).\label{Phi'(r)Mod}%
\end{equation}
For the energy density profiles we are going to examine, it will be sufficient
to consider $\alpha=1$. Then Eq.$\left(  \ref{or(r)mod}\right)  $ reduces to%
\begin{equation}
\omega_{r}\left(  r\right)  =-\frac{r_{0}b(r)}{r^{2}b^{\prime}(r)}%
,\label{or(r)Ma1}%
\end{equation}
while Eq.$\left(  \ref{Phi'(r)Mod}\right)  $ reduces to%
\begin{equation}
\Phi^{\prime}(r)=\frac{b(r)\left(  r-r_{0}\right)  }{2r^{2}\left(
r-b(r)\right)  }.\label{Phi'(r)Ma1}%
\end{equation}
Finally, plugging Eq.$\left(  \ref{or(r)Ma1}\right)  $ and Eq.$\left(
\ref{Phi'(r)Ma1}\right)  $ into Eq.$\left(  \ref{Tmn}\right)  $, one finds%
\begin{equation}
\omega_{t}\!\left(  r\right)  =\frac{\left(  r^{2}\left(  r+r_{0}\right)
b\!\left(  r\right)  -2r^{3}r_{0}\right)  b\!^{\prime}\left(  r\right)
-r_{0}\left(  \left(  r_{0}-5r\right)  b\!\left(  r\right)  +4r_{0}%
r^{2}\right)  b\!\left(  r\right)  }{4b\!^{\prime}\left(  r\right)
r^{3}\left(  r-b\!\left(  r\right)  \right)  },\label{ot(r)Ma1}%
\end{equation}
where we have also used the additional EoS%
\begin{equation}
p_{t}(r)=\omega_{t}\left(  r\right)  \rho(r).\label{EoS2}%
\end{equation}
Now we have all the elements to examine the different energy density profiles.
We begin to examine the BH HDE. As we will see, this is the only example under
examination which does not need a modification of the EoS.

\textbf{Remark }The modification $\left(  \ref{or(r)mod}\right)  $ and its
special case $\left(  \ref{or(r)Ma1}\right)  $ can be taken under
consideration because $p_{r}(r)$ and $p_{t}(r)$ are not fixed. Only the energy
density has a known profile. Indeed for the Casimir wormhole, this procedure
cannot be applied, because the original SET is known.

\section{Bekenstein-Hawking (BH) HDE}
\label{ch4}
As a potential source for TW, we consider the Bekenstein-Hawking HDE density,
whose profile is described by Eq.$\left(  \ref{BH}\right)  $, which here we
report \cite{Manoharan:2022qll}
\begin{equation}
\rho_{1}\left(  r\right)  =\frac{C\pi}{r^{2}}.
\end{equation}
$C$ is positive with dimensions of $\left[  L^{-2}\right]  $. Plugging the
energy density $\left(  \ref{BH}\right)  $ into Eq.$\left(  \ref{rho}\right)
$, one finds%
\begin{equation}
b(r)=r_{0}+\beta\left(  r-r_{0}\right)  =r_{0}\left(  1-\beta\right)  +\beta
r,\label{b(r)1}%
\end{equation}
where $\beta=\kappa C\pi$. For $r\rightarrow\infty$, $b(r)\rightarrow\infty$.
The flare-out condition demands that%
\begin{equation}
b^{\prime}(r_{0})<1,
\end{equation}
namely $\beta<1$. This means that $b(r)$ never vanishes. The line element
$\left(  \ref{ds}\right)  $ can be written in the following way%
\begin{equation}
ds^{2}=-e^{2\Phi(r)}dt^{2}+\frac{dr^{2}}{\left(  1-\beta\right)  \left(
1-\frac{r_{0}}{r}\right)  }+r^{2}\left(  d\theta^{2}+\sin^{2}\theta d\phi
^{2}\right)  ,\label{ds21}%
\end{equation}
with $\beta\neq1$. This is in agreement with what has been found in
Ref.\cite{CLR}, where the ZTF case has been discussed leading to $\Phi(r)=0$,
even if nothing has been said about a possible EoS allowing such a choice.

\subsection{Homogeneous Radial EoS}

\label{HREoS}In this subsection, we adopt the strategy used in Refs.\cite{EPJC1, Garattini:2021kca} and consider Eq.$\left(  \ref{pr}\right)  $%
\begin{equation}
\left[  \frac{2}{r}\left(  \left(  1-\beta\right)  \left(  1-\frac{r_{0}}%
{r}\right)  \right)  \Phi^{\prime}\left(  r\right)  -\left(  1-\beta\right)
\frac{r_{0}}{r^{3}}-\frac{\beta}{r^{2}}\right]  =\kappa p_{r}(r),
\end{equation}
where we have used Eq.$\left(  \ref{b(r)1}\right)  $ and the EoS
$p_{r}(r)=\omega_{r}\rho(r)$. Solving with respect to $\Phi\left(  r\right)
$, one gets%
\begin{equation}
\Phi\!\left(  r\right)  =\frac{\ln\!\left(  r-r_{0}\right)  }{2\left(
1-\beta\right)  }\left(  1+\beta\omega_{r}\right)  -\frac{\ln\!\left(
r\right)  }{2}+K.\label{Phi(r)}%
\end{equation}
When $r\rightarrow r_{0}$, a horizon is present. It is straightforward to see
that if we impose that%
\begin{equation}
\omega_{r}=-\frac{1}{\beta},\label{omega}%
\end{equation}
then the redshift function is regular for $r=r_{0}$ and one gets%
\begin{equation}
\Phi\!\left(  r\right)  =\frac{1}{2}\ln\!\left(  r_{0}/r\right)  ,
\end{equation}
where we have assumed that $\Phi\!\left(  r_{0}\right)  =0$. However, such a
choice is not complete, because the behavior of $\Phi\!\left(  r\right)  $ for
$r\rightarrow+\infty$ has not been determined yet. From Eq.$\left(
\ref{Phi(r)}\right)  $, one gets%
\begin{equation}
\Phi\!\left(  r\right)  \underset{r\rightarrow\infty}{\simeq}\frac
{\ln\!\left(  r\right)  }{2}\left[  \frac{\beta\left(  1+\omega_{r}\right)
}{1-\beta}\right]  +K,
\end{equation}
that it means that%
\begin{equation}
\omega_{r}=-1\label{omegai}%
\end{equation}
to have a finite result. The line element $\left(  \ref{ds21}\right)  $ becomes%
\begin{align}
ds^{2} &  =-\frac{r_{0}}{r}dt^{2}+\frac{dr^{2}}{\left(  1-\beta\right)
\left(  1-\frac{r_{0}}{r}\right)  }+r^{2}\left(  d\theta^{2}+\sin^{2}\theta
d\phi^{2}\right)  \qquad r\rightarrow r_{0},\label{ds21a}\\
& \nonumber\\
ds^{2} &  =-dt^{2}+\frac{dr^{2}}{1-\beta}+r^{2}\left(  d\theta^{2}+\sin
^{2}\theta d\phi^{2}\right)  \qquad\qquad\qquad\qquad r\rightarrow
\infty.\label{ds21aa}%
\end{align}
Note that for $r\rightarrow\infty$, we are in presence of a \textit{Global
Monopole} \cite{EPJC1}. Since $1<\beta$, we have an excess of the solid angle
for the line element $\left(  \ref{ds21aa}\right)  $ which can be cast in the
following way%
\begin{equation}
ds^{2}=-dt^{2}+d\tilde{r}^{2}+\left(  1-\beta\right)  \tilde{r}^{2}\left(
d\theta^{2}+\sin^{2}\theta d\phi^{2}\right)  .
\end{equation}
Finally we consider Eq.$\left(  \ref{pt}\right)  $ which reduces to%
\begin{align}
p_{t}(r) &  =\frac{1-\beta}{4\kappa r^{2}}\qquad r\rightarrow r_{0}%
,\label{ptBHa}\\
& \nonumber\\
p_{t}(r) &  =\frac{\left(  1-\beta\right)  r_{0}}{2\kappa r^{3}}\qquad
r\rightarrow\infty.\label{ptBHb}%
\end{align}
To summarize, the Stress Energy Tensor (SET) becomes%
\begin{align}
T_{\mu\nu} &  =\frac{1}{\kappa\,{r}^{2}}\left[  diag\left(  \beta
,-1,\frac{1-\beta}{4},\frac{1-\beta}{4}\right)  \right]  \qquad r\rightarrow
r_{0},\label{SETa}\\
& \nonumber\\
T_{\mu\nu} &  =\frac{1}{\kappa\,{r}^{2}}\left[  diag\left(  \beta,-\beta
,\frac{\left(  1-\beta\right)  r_{0}}{2r},\frac{\left(  1-\beta\right)  r_{0}%
}{2r}\right)  \right]  \qquad r\rightarrow\infty.
\end{align}
Note that this SET cannot be traceless, because this should imply $\beta=-1/3$
for $r\rightarrow r_{0}$. This is not possible since $C>0$. On the other hand,
for $r\rightarrow\infty$, $\beta=0$ which is inconsistent with the original
energy density profile. Since in the homogeneous case, $\omega_{r}$ must
assume two distinct values in two different spatial regions, we are going to
examine the inhomogeneous EoS to see if there exists a unique choice for the EoS.

\subsection{Inhomogeneous Radial EoS}

\label{IREoS}If we consider the relationship $\left(  \ref{or(r)}\right)  $,
we can set $\Phi\!\left(  r\right)  =0$ everywhere and%
\begin{equation}
\omega_{r}\left(  r\right)  =-\frac{r_{0}\left(  1-\beta\right)  +\beta
r}{\beta r}.
\end{equation}
We observe that%
\begin{equation}
\omega_{r}\left(  r_{0}\right)  =-\frac{1}{\beta}\qquad\mathrm{and}%
\qquad\omega_{r}\left(  r\right)  \underset{r\rightarrow\infty}{=}-1,
\end{equation}
which is consistent with what has been found in section \ref{HREoS}. Note
that, since $0<\beta<1$, $\omega_{r}\left(  r\right)  $ tells us that we are
in presence of phantom energy. The line
element now becomes%
\begin{equation}
ds^{2}=-dt^{2}+\frac{dr^{2}}{\left(  1-\beta\right)  \left(  1-\frac{r_{0}}%
{r}\right)  }+r^{2}\left(  d\theta^{2}+\sin^{2}\theta d\phi^{2}\right)  ,
\label{ds21b}%
\end{equation}
which is the same one which has been found in Ref.\cite{CLR}. With the help of
the shape function $\left(  \ref{form}\right)  $ and $\omega_{r}\left(
r\right)  $, it is possible to write the SET in its general representation,
namely%
\begin{gather}
T_{\mu\nu}=\frac{r_{0}}{\kappa r^{3}}diag\left(  -\frac{1}{\omega_{r}\left(
r\right)  },-1,\frac{1}{2\omega_{r}\left(  r\right)  }+\frac{1}{2},\frac
{1}{2\omega_{r}\left(  r\right)  }+\frac{1}{2}\right)  \exp\left[
-\int_{r_{0}}^{r}\,\frac{d\bar{r}}{\omega_{r}\left(  \bar{r}\right)  \bar{r}%
}\right] \nonumber\\
=-\frac{b(r)}{\kappa r^{3}\omega_{r}\left(  r\right)  }diag\left(
1,\omega_{r}\left(  r\right)  ,-\frac{1}{2}-\frac{\omega_{r}\left(  r\right)
}{2},-\frac{1}{2}-\frac{\omega_{r}\left(  r\right)  }{2}\right) \nonumber\\
=\rho\left(  r\right)  diag\left(  1,\omega_{r}\left(  r\right)  ,-\frac{1}%
{2}-\frac{\omega_{r}\left(  r\right)  }{2},-\frac{1}{2}-\frac{\omega
_{r}\left(  r\right)  }{2}\right)  . \label{SET1}%
\end{gather}

\subsection{SET Conservation}

With the help of the Equations of State $\left(  \ref{EoS1}\right)  $ and
$\left(  \ref{EoS2}\right)  $, the SET conservation described by Eq.$\left(
\ref{Tmn}\right)  $, becomes%
\begin{equation}
\frac{d}{dr}\left(  \omega_{r}\left(  r\right)  \rho(r)\right)  =\frac{2}%
{r}\left(  \omega_{t}\left(  r\right)  -\omega_{r}\left(  r\right)  \right)
\rho(r)-\left(  1+\omega_{r}\left(  r\right)  \right)  \rho\left(  r\right)
\Phi^{\prime}(r).
\end{equation}
Isolating $\Phi^{\prime}\left(  r\right)  $, one gets%
\begin{equation}
\Phi^{\prime}(r)=-\frac{\omega_{r}^{\prime}\left(  r\right)  }{1+\omega
_{r}\left(  r\right)  }-\frac{\omega_{r}\left(  r\right)  \rho^{\prime}\left(
r\right)  }{\left(  1+\omega_{r}\left(  r\right)  \right)  \rho\left(
r\right)  }+\frac{2}{r}\frac{\left(  \omega_{t}\left(  r\right)  -\omega
_{r}\left(  r\right)  \right)  }{1+\omega_{r}\left(  r\right)  }%
.\label{TmnEoS}%
\end{equation}
In the case of BH HDE, Eq.$\left(  \ref{TmnEoS}\right)  $ leads to the
following equation for the redshift function%
\begin{equation}
\Phi\!^{\prime}\left(  r\right)  =\frac{-\omega_{r}^{\prime}\!\left(
r\right)  r+2\omega_{t}\!\left(  r\right)  }{r\left(  \omega_{r}\!\left(
r\right)  +1\right)  },\label{PhiSETBH}%
\end{equation}
where the profile $\left(  \ref{BH}\right)  $ has been used. Differently from
the other profiles we will investigate, this one can offer solutions without
approximations. Before doing this we need to verify the consistency with
Eq.$\left(  \ref{or(r)}\right)  $. To this purpose, Eq.$\left(  \ref{PhiSETBH}%
\right)  $ can be rearranged to give%
\begin{equation}
\frac{d}{dr}\left(  \Phi\left(  r\right)  +\ln\left(  \omega_{r}\!\left(
r\right)  +1\right)  \right)  =\frac{2\omega_{t}\!\left(  r\right)  }{r\left(
\omega_{r}\!\left(  r\right)  +1\right)  }.
\end{equation}
If we assume that%
\begin{equation}
\omega_{t}\!\left(  r\right)  =-\frac{\omega_{r}\!\left(  r\right)  +1}%
{2},\label{otor}%
\end{equation}
then we find%
\begin{equation}
\frac{d}{dr}\left(  \Phi\left(  r\right)  +\ln\left(  \left\vert \omega
_{r}\!\left(  r\right)  +1\right\vert \right)  +\ln\left(  r\right)  \right)
=0,
\end{equation}
namely%
\begin{equation}
\Phi\left(  r\right)  +\ln\left(  r\left(  \left\vert \omega_{r}\!\left(
r\right)  +1\right\vert \right)  \right)  =K.
\end{equation}
The assumption in Eq.$\left(  \ref{otor}\right)  $ is suggested by the SET in
Eq.$\left(  \ref{SET1}\right)  $. The consistency with Eq.$\left(
\ref{or(r)}\right)  $ is guaranteed if $\Phi\left(  r\right)  =0$. On the
other hand to obtain other redshift profiles, we can plug Eq.$\left(
\ref{PhiSETBH}\right)  $ into Eq.$\left(  \ref{pr}\right)  $ and we find%
\begin{equation}
2\left(  1-\frac{b(r)}{r}\right)  \frac{-\omega_{r}^{\prime}\!\left(
r\right)  r+2\omega_{t}\!\left(  r\right)  }{r^{2}\left(  \omega_{r}\!\left(
r\right)  +1\right)  }-\frac{b(r)}{r^{3}}-\frac{\omega_{r}\!\left(  r\right)
b^{\prime}(r)}{r^{2}}=0,\label{prBH}%
\end{equation}
where we have used the EoS $\left(  \ref{EoS1}\right)  $. With the help of the
shape function $\left(  \ref{b(r)1}\right)  $ and the relationship $\left(
\ref{otor}\right)  $, Eq.$\left(  \ref{prBH}\right)  $ becomes%
\begin{equation}
\frac{d}{dr}\omega_{r}\!\left(  r\right)  =-\frac{\left(  \omega_{r}\!\left(
r\right)  \beta r-\left(  \beta-2\right)  r-\left(  1-\beta\right)
r_{0}\right)  \left(  \omega_{r}\!\left(  r\right)  +1\right)  }{2r\left(
1-\beta\right)  \left(  r-r_{0}\right)  },
\end{equation}
whose solution is%
\begin{equation}
\omega_{r}\!\left(  r\right)  =-\frac{\sqrt{r-r_{0}}\sqrt{r}+\left(  \left(
r-r_{0}\right)  \beta+r_{0}\right)  C_{1}}{\sqrt{r}\left(  C_{1}\beta\sqrt
{r}+\sqrt{r-r_{0}}\right)  },\label{orBH}%
\end{equation}
where $C_{1}$ is an arbitrary constant. $\omega_{r}\!\left(  r\right)  $ is
such that%
\begin{equation}
\omega_{r}\!\left(  r_{0}\right)  =-\frac{1}{\beta}\qquad and\qquad\omega
_{r}\!\left(  r\right)  \underset{r\rightarrow\infty}{\longrightarrow}-1.
\end{equation}
As a consequence, from the relationship $\left(  \ref{otor}\right)  $,%
\begin{equation}
\omega_{t}\!\left(  r_{0}\right)  =\frac{1-\beta}{2\beta}\qquad and\qquad
\omega_{t}\!\left(  r\right)  \underset{r\rightarrow\infty}{\longrightarrow}0.
\end{equation}
Note that the integration constant $C_{1}$ is not determined. Plugging
Eq.$\left(  \ref{orBH}\right)  $ into Eq.$\left(  \ref{PhiSETBH}\right)  $,
one finds%
\begin{equation}
\frac{d}{dr}\Phi\!\left(  r\right)  =\frac{r_{0}}{2r\left(  C_{1}\beta\sqrt
{r}+\sqrt{r-r_{0}}\right)  \sqrt{r-r_{0}}}.\label{PhiSETBH1}%
\end{equation}
It is immediate to see that for $C_{1}=0$, Eq.$\left(  \ref{PhiSETBH1}\right)
$ develops a horizon. Therefore this option will be discarded. The general
solution of Eq.$\left(  \ref{PhiSETBH1}\right)  $ is%
\begin{equation}
\Phi\!\left(  r\right)  =\frac{1}{2}\ln\!\left(  \frac{C_{1}\beta\sqrt
{r}+\sqrt{r-r_{0}}}{C_{1}\beta\sqrt{r}-\sqrt{r-r_{0}}}\right)  +\frac{1}{2}%
\ln\!\left(  \frac{\left(  C_{1}^{2}\beta^{2}-1\right)  r+r_{0}}{C_{1}%
^{2}\beta^{2}r}\right)  ,
\end{equation}
where we have assumed that $\Phi\!\left(  r_{0}\right)  =0$.



\section{Moradpour energy density}

\label{ch5}In this section, we are going to examine the following profile \cite{Manoharan:2022qll}
\begin{equation}
\rho_{2}\left(  r\right)  =\frac{C}{4\pi r^{2}\left(  \pi\lambda
r^{2}+1\right)  }=\frac{C}{4\pi}\left(  \frac{1}{r^{2}}-\frac{\pi\lambda}%
{\pi\lambda r^{2}+1}\right)  .
\end{equation}
$\rho_{2}\left(  r\right)  $ vanishes for $r\rightarrow\infty$, as well as for
$\lambda\rightarrow\infty$. For this asymptotic cases $b(r)=r_{0}$. On the
other hand when $\lambda=0$, we obtain the energy density described by
Eq.$\left(  \ref{BH}\right)  $ with an additional $4\pi$ term at the
denominator. Note that $C$ and $\lambda$ have dimensions $\left[
L^{-2}\right]  $. Plugging $\rho_{2}\left(  r\right)  $ into the first EFE
$\left(  \ref{rho}\right)  $, we find%
\begin{equation}
b(r)=r_{0}+\frac{\kappa C}{4\pi\sqrt{\pi\lambda}}\left(  \tan^{-1}\left(
\sqrt{\pi\lambda}r\right)  -\tan^{-1}\left(  \sqrt{\pi\lambda}r_{0}\right)
\right)  ,\label{b(r)2}%
\end{equation}
which is always positive. This shape function is such that%
\begin{equation}
b(r)=\underset{r\rightarrow\infty}{\rightarrow}r_{0}+\frac{\kappa C}{4\pi
\sqrt{\pi\lambda}}\left(  \pi/2-\tan^{-1}\left(  \sqrt{\pi\lambda}%
\,r_{0}\right)  \right)  =b_{M,\infty},\label{b(r)2a}%
\end{equation}
which reduces to the value $r_{0}$ when $\lambda\rightarrow\infty$, as it
should be. Since $\lambda\geq0$, the flare-out condition, represented by
\begin{equation}
b^{\prime}(r_{0})=\frac{\kappa C}{4\pi\left(  \pi\lambda r_{0}^{2}+1\right)
}<1,\label{b'(r0)M}%
\end{equation}
is always satisfied. To gain enough information on the $\Phi\left(  r\right)
$, we examine the original inhomogeneous EoS to see if a modification is
necessary. Such an EoS, if satisfied, allows us to impose ZTF and set
$\Phi\!\left(  r\right)  =0$ everywhere. To this purpose, we need to compute
Eq.$\left(  \ref{or(r)}\right)  $ which is represented by
\begin{equation}
\omega_{r}\left(  r\right)  =-\frac{1+\pi\lambda r^{2}}{\kappa C\sqrt
{\pi\lambda}r}\left(  4\pi\sqrt{\pi\lambda}r_{0}+\kappa C\left(  \tan
^{-1}\left(  \sqrt{\pi\lambda}r\right)  -\tan^{-1}\left(  \sqrt{\pi\lambda
}r_{0}\right)  \right)  \right)  .\label{or(r)M}%
\end{equation}
On the throat we find%
\begin{equation}
\omega_{r}\left(  r_{0}\right)  =-\frac{4\pi}{\kappa C}\left(  1+\pi\lambda
r_{0}^{2}\right)  <0,\label{or(r)Mr0}%
\end{equation}
while for $r\rightarrow\infty$, one gets%
\begin{equation}
\omega_{r}\left(  r\right)  \underset{r\rightarrow\infty}{\simeq}=-\left(
\frac{8\pi\sqrt{\pi\lambda}r_{0}+\kappa C\left(  \pi-2\tan^{-1}\left(
\sqrt{\pi\lambda}r_{0}\right)  \right)  }{2\kappa C}\right)  \sqrt{\pi\lambda
}r\rightarrow\infty.\label{or(r)L}%
\end{equation}
Since the quantity inside the round brackets never vanishes, $\omega
_{r}\left(  r\right)  $ diverges for $r\rightarrow\infty$. Therefore the ZTF
cannot be imposed. However, we can use the modification $\left(
\ref{or(r)Ma1}\right)  $ to obtain%
\begin{equation}
\omega_{r}\left(  r\right)  =-\frac{r_{0}\left(  1+\pi\lambda r^{2}\right)
}{\kappa C\sqrt{\pi\lambda}r^{2}}\left(  4\pi\sqrt{\pi\lambda}r_{0}+\kappa
C\left(  \tan^{-1}\left(  \sqrt{\pi\lambda}r\right)  -\tan^{-1}\left(
\sqrt{\pi\lambda}r_{0}\right)  \right)  \right)  .\label{or(r)MM1}%
\end{equation}
and this time, for $r\rightarrow\infty$, one finds%
\begin{equation}
\omega_{r}\left(  r\right)  \underset{r\rightarrow\infty}{\simeq}=-\frac
{r_{0}\sqrt{\pi\lambda}}{2\kappa C}\left(  8\pi\sqrt{\pi\lambda}r_{0}+\kappa
C\left(  \pi-2\tan^{-1}\left(  \sqrt{\pi\lambda}r_{0}\right)  \right)
\right)  .
\end{equation}
The redshift function is described by Eq.$\left(  \ref{Phi'(r)Ma1}\right)  $
which, in this particular case, becomes%
\begin{equation}
\Phi^{\prime}(r)\simeq\frac{r_{0}}{2\left(  1-B\right)  r^{2}},
\end{equation}
where we have used the approximation $\left(  \ref{b(r)t}\right)  $ and where
$B$ is represented by Eq.$\left(  \ref{b'(r0)M}\right)  $. If we assume that
$\Phi(r_{0})=0$, then one finds%
\begin{equation}
\Phi(r)\simeq\frac{1}{2\left(  1-B\right)  }\left(  1-\frac{r_{0}}{r}\right)
.\label{Phi(r)MT}%
\end{equation}
On the other hand, when $r\rightarrow\infty$, we can use the asymptotic
behavior of $b\left(  r\right)  $ described in $\left(  \ref{b(r)2a}\right)  $
to obtain%
\begin{equation}
\Phi^{\prime}(r)\simeq\frac{b_{M,\infty}}{2r^{2}}\qquad\Longrightarrow
\qquad\Phi(r)\simeq-\frac{b_{M,\infty}}{2r}.\label{Phi(r)MA}%
\end{equation}
To have consistency between $\Phi(r)$, the SET equation and the third EFE, we
will use Eq.$\left(  \ref{ot(r)Ma1}\right)  $. Close to the throat, we can
write%
\begin{equation}
\omega_{t}\!\left(  r\right)  =\frac{\left(  -r^{3}+4r^{2}r_{0}-6rr_{0}%
^{2}+r_{0}^{3}\right)  B^{2}+\left(  -3r^{2}r_{0}+10rr_{0}^{2}-2r_{0}%
^{3}\right)  B-4rr_{0}^{2}+r_{0}^{3}}{4B\,r^{3}\left(  B-1\right)  },
\end{equation}
which implies%
\begin{equation}
\omega_{t}\!\left(  r_{0}\right)  =\frac{3-2B}{4B},\label{ot(r0)M}%
\end{equation}
where we have used the approximation $\left(  \ref{b(r)t}\right)  $. Note
that, since $0<B<1$, $\omega_{t}\!\left(  r_{0}\right)  >0$. On the other
hand, when $r\rightarrow\infty$, one gets
\begin{equation}
\omega_{t}\!\left(  r\right)  \simeq r_{0}\left(  \frac{\pi\kappa C\sqrt
{\pi\lambda}-2\sqrt{\pi\lambda}\kappa C\arctan\!\left(  \sqrt{\pi\lambda}%
r_{0}\right)  +8\pi^{2}r_{0}\lambda}{2\kappa C}\right)  +O\left(  \frac
{1}{r^{3}}\right)  .\label{ot(r)ML}%
\end{equation}
The same results hold also for the third EFE as it should be. We can observe
that, although the original $\omega_{r}\left(  r\right)  $ of Eq.$\left(
\ref{or(r)M}\right)  $ is divergent when $r\rightarrow\infty$, the radial
pressure given by Eq.$\left(  \ref{EoS1}\right)  $ is not. This is due to the
action of the energy density that decreases like $1/r^{2}$ leading to a radial
pressure that decreases like $1/r$. The modification $\left(  \ref{or(r)Ma1}%
\right)  $ allows not only to have a convergent $\omega_{r}\left(  r\right)
$, but also a radial pressure that goes at infinity like $1/r^{2}$. Regarding
the transverse pressure, we can see that Eq.$\left(  \ref{EoS2}\right)  $ and
the approximation $\left(  \ref{ot(r)ML}\right)  $ tell us that $p_{t}\left(
r\right)  \simeq1/r^{2}$ when $r\rightarrow\infty$.

\section{Standard Renyi HDE}

\label{ch6}Here the Renyi HDE density from the CKN bound takes the form \cite{Manoharan:2022qll}
\begin{equation}
\rho_{3}\left(  r\right)  =\frac{C}{\lambda r^{4}}\ln\left(  1+\pi\lambda
r^{2}\right)  .\label{rho3}%
\end{equation}
Note that $C$ and $\lambda$ have dimensions $\left[  L^{-2}\right]  $ and are
positive. $\rho_{3}\left(  r\right)  $ vanishes for $r\rightarrow\infty$, as
well as for $\lambda\rightarrow\infty$. On the other hand for $\lambda
\rightarrow0,$ $\rho_{3}\left(  r\right)  \rightarrow$ $\rho_{1}\left(
r\right)  $ described by Eq.$\left(  \ref{BH}\right)  $. Plugging Eq.$\left(
\ref{rho3}\right)  $ into Eq.$\left(  \ref{rho}\right)  $, one finds%
\begin{align}
b(r) &  =r_{0}+\frac{\kappa C}{\lambda}\left(  \frac{\ln\left(  \pi\lambda
r_{0}^{2}+1\right)  }{r_{0}}-\frac{\ln\left(  \pi\lambda r^{2}+1\right)  }%
{r}\right)  \nonumber\\
&  +\frac{2\kappa C\sqrt{\pi}}{\sqrt{\lambda}}\left(  \tan^{-1}\left(
\sqrt{\pi\lambda}r\right)  -\tan^{-1}\left(  \sqrt{\pi\lambda}r_{0}\right)
\right)  .\label{b(r)3}%
\end{align}
Note that even for this profile when $\lambda\rightarrow\infty$, $b(r)=r_{0}$.
This shape function is such that%
\begin{equation}
b(r)\underset{r\rightarrow\infty}{\rightarrow}r_{0}+\frac{\pi^{3/2}C\kappa
}{\sqrt{\lambda}}+\frac{C\kappa\ln\left(  \pi\lambda r_{0}^{2}+1\right)
}{\lambda r_{0}}-\frac{2\sqrt{\pi}C\kappa\tan^{-1}\left(  \sqrt{\pi\lambda
}r_{0}\right)  }{\sqrt{\lambda}}=b_{R,\infty}.
\end{equation}
Since $\lambda>0$, the flare-out condition, described by the following
inequality%
\begin{equation}
b^{\prime}(r_{0})=\frac{C\kappa\ln\left(  \pi\lambda r_{0}^{2}+1\right)
}{\lambda r_{0}^{2}}<1,\label{b'(r0)R}%
\end{equation}
is always satisfied. Even for this profile, we will try to see if it is
possible to set $\Phi\!\left(  r\right)  =0$ everywhere by means of the
relationship $\left(  \ref{or(r)}\right)$. One gets%
\begin{equation}
\omega_{r}\left(  r\right)  =1+\frac{r}{\ln\!\left(  \pi\lambda r^{2}%
+1\right)  r_{0}}\left(  2r_{0}\sqrt{\lambda\pi}\,\arctan\!\left(  \sqrt
{\pi\lambda}r_{0}\right)  -\frac{\lambda r_{0}^{2}}{\kappa C}-2r_{0}%
\sqrt{\lambda\pi}\arctan\!\left(  \sqrt{\pi\lambda}r\right)  -\ln\!\left(
\pi\lambda r_{0}^{2}+1\right)  \right)  .\label{or(r)R}%
\end{equation}
This means that, even in this case,we find a divergent inhomogeneous
$\omega_{r}\left(  r\right)  $. Of course, we can impose ZTF, but at the price
of having a divergent inhomogeneous $\omega_{r}\left(  r\right)  $ for large
values of $r$. Therefore the ZTF case will be discarded like for the Moradpour
profile. However, following the same procedure of the previous section, we can
modify the form of $\omega_{r}\left(  r\right)  $ in such a way to compensate
the divergent behavior. Indeed, from Eq.$\left(  \ref{or(r)R}\right)  $, one
finds that $\omega_{r}\left(  r\right)  \sim r$ when $r\rightarrow\infty$.
Thus, if we adopt Eq.$\left(  \ref{ot(r)Ma1}\right)  $ also for the Renyi
profile, we find%
\begin{equation}
\omega_{r}\left(  r\right)  =\frac{r_{0}}{r}+\frac{1}{\ln\!\left(  \pi\lambda
r^{2}+1\right)  }\left(  2r_{0}\sqrt{\lambda\pi}\,\arctan\!\left(  \sqrt
{\pi\lambda}r_{0}\right)  -\frac{\lambda r_{0}^{2}}{\kappa C}-2r_{0}%
\sqrt{\lambda\pi}\arctan\!\left(  \sqrt{\pi\lambda}r\right)  -\ln\!\left(
\pi\lambda r_{0}^{2}+1\right)  \right)  .\label{or(r)R1}%
\end{equation}
Now $\omega_{r}\left(  r\right)  \rightarrow0$ when $r\rightarrow\infty$ and,
on the throat, we get the same expression of Eq.$\left(  \ref{or(r)R}\right)
$. It is clear that the redshift function obeys the differential equation
$\left(  \ref{Phi'(r)Ma1}\right)  $ whose solution close to the throat is
represented by Eq.$\left(  \ref{Phi(r)MT}\right)  $. Only the value of
$B=b^{\prime}(r_{0})$ represented by Eq.$\left(  \ref{b'(r0)R}\right)  $ is
different as it should be. On the other hand, when $r\rightarrow\infty$, we
can use the asymptotic expression of $b\left(  r\right)  $ leading to the same
analytic form of $\left(  \ref{Phi(r)MA}\right)  $, but with $b_{M,\infty}$
replaced by $b_{R,\infty}$. It is easy to check that also for the Renyi
profile, $\omega_{t}\!\left(  r\right)  $ assumes the same analytic expression
of Eq.$\left(  \ref{ot(r)Ma1}\right)  $. This means that on the throat we will
obtain the same value described in Eq.$\left(  \ref{ot(r0)M}\right)  $. Only
the value of $B$ will be different as it should be. On the other hand, when
$r\rightarrow\infty$, one gets%
\begin{equation}
\omega_{t}\left(  r\right)  \simeq\frac{C\kappa\pi\sqrt{\pi\lambda}%
r_{0}-2C\kappa\sqrt{\pi\lambda}r_{0}\arctan\!\left(  \sqrt{\pi\lambda}%
r_{0}\right)  +\kappa C\ln\!\left(  \pi\lambda r_{0}^{2}+1\right)  +\lambda
r_{0}^{2}}{C\kappa\ln\!\left(  \pi\lambda r^{2}\right)  }.\label{ot(r)RL}%
\end{equation}
The same results hold also for the third EFE as it should be. Even for the
Renyi profile, we can observe that, although the original $\omega_{r}\left(
r\right)  $ of Eq.$\left(  \ref{or(r)R}\right)  $ is divergent when
$r\rightarrow\infty$, the radial pressure given by Eq.$\left(  \ref{EoS1}%
\right)  $ is not. The reason is the same of the previous section: the
behavior of the energy density when $r\rightarrow\infty$ is $\rho
(r)\sim1/r^{2}$ leading to a radial pressure that decreases like $1/r$. The
modification $\left(  \ref{or(r)R1}\right)  $ allows not only to have a
convergent $\omega_{r}\left(  r\right)  $, but also a radial pressure that
goes at infinity like $1/r^{2}$. Regarding the transverse pressure, we can see
that Eq.$\left(  \ref{EoS2}\right)  $ and the approximation $\left(
\ref{ot(r)ML}\right)  $ tell us that $p_{t}\left(  r\right)  \simeq1/r^{2}$
when $r\rightarrow\infty$.

\section{Mixed Energy Density}

\label{ch7}In this section we consider a combination of the form
\begin{equation}
\rho_{4}\left(  r\right)  =\frac{3C_{M}^{2}}{8\pi^{2}}\left[  \frac{\pi}%
{r^{2}}-\pi^{2}\lambda\ln\left(  1+\frac{1}{\pi\lambda r^{2}}\right)  \right]
.\label{rho4}%
\end{equation}
Note that, $C_{M}$ has dimensions $\left[  L^{-1}\right]  $ while $\lambda$
has dimensions $\left[  L^{-2}\right]  $ and both are positive. For
$r\rightarrow\infty$ and $\lambda\rightarrow\infty$, $\rho_{4}\left(
r\right)  \rightarrow0$. Therefore for $\lambda\rightarrow\infty$,
$b(r)=r_{0}$ represents a solution. For $\lambda\rightarrow0$, the energy
density reduces to the Bekenstein-Hawking (BH) HDE profile of section
\ref{ch4} for an appropriate choice of the constant $C_{M}$. Plugging
Eq.$\left(  \ref{rho4}\right)  $ into Eq.$\left(  \ref{rho}\right)  $, one
finds%
\begin{equation}
b(r)=r_{0}+\frac{\kappa C_{M}^{2}}{8}\left[  \pi r_{0}^{3}\lambda\ln\!\left(
\frac{1+\pi\lambda r_{0}^{2}}{\lambda\pi r_{0}^{2}}\right)  -\frac
{2\arctan\!\left(  \sqrt{\pi\lambda}r_{0}\right)  }{\sqrt{\pi\lambda}}%
+\frac{2\arctan\!\left(  \sqrt{\pi\lambda}r\right)  }{\sqrt{\pi\lambda}}%
-r^{3}\lambda\pi\left(  \ln\!\left(  \frac{\pi\lambda\,r^{2}+1}{\lambda\pi
r^{2}}\right)  \right)  +\left(  r-r_{0}\right)  \right]  .
\end{equation}
This shape function is such that%
\begin{equation}
b(r)\underset{r\rightarrow\infty}{\rightarrow}r_{0}+\frac{\kappa C_{M}^{2}}%
{8}\left[  \pi r_{0}^{3}\lambda\ln\!\left(  \frac{1+\pi\lambda r_{0}^{2}%
}{\lambda\pi r_{0}^{2}}\right)  -\frac{2\arctan\!\left(  \sqrt{\pi\lambda
}r_{0}\right)  }{\sqrt{\pi\lambda}}+\frac{\!\sqrt{\pi}}{\sqrt{\lambda}}%
+r_{0}\right]  =b_{Mix,\infty},
\end{equation}
while the flare-out condition is described by the following inequality%
\begin{equation}
b^{\prime}(r_{0})=\frac{3\kappa C_{M}^{2}}{8}\left[  1-r_{0}^{2}\lambda
\pi\left(  \ln\!\left(  \frac{\pi\lambda\,r_{0}^{2}+1}{\lambda\pi r_{0}^{2}%
}\right)  \right)  \right]  <1.\label{b'(r0)Mix}%
\end{equation}
It is easy to see that the previous inequality is always satisfied. Now, we
need to know if the ZTF can be imposed. To this purpose, we compute
$\omega_{r}\left(  r\right)  $, like in Eq.$\left(  \ref{or(r)}\right)  $.
Since the expression
\begin{equation}
\omega_{r}\left(  r\right)  =-\frac{\left\{  r_{0}+\frac{\kappa C_{M}^{2}}%
{8}\left[  \pi r_{0}^{3}\lambda\ln\!\left(  \frac{1+\pi\lambda r_{0}^{2}%
}{\lambda\pi r_{0}^{2}}\right)  -\frac{2\arctan\!\left(  \sqrt{\pi\lambda
}r_{0}\right)  }{\sqrt{\pi\lambda}}+\frac{2\arctan\!\left(  \sqrt{\pi\lambda
}r\right)  }{\sqrt{\pi\lambda}}-r^{3}\lambda\pi\left(  \ln\!\left(  \frac
{\pi\lambda\,r^{2}+1}{\lambda\pi r^{2}}\right)  \right)  +\left(
r-r_{0}\right)  \right]  \right\}  }{\frac{3\kappa C_{M}^{2}}{8}r\left[
1-r^{2}\lambda\pi\left(  \ln\!\left(  \frac{\pi\lambda\,r^{2}+1}{\lambda\pi
r^{2}}\right)  \right)  \right]  }.
\end{equation}
On the throat, we find%
\begin{equation}
\omega_{r}\left(  r_{0}\right)  =-\frac{8}{3\kappa C_{M}^{2}\left[
1-r_{0}^{2}\lambda\pi\left(  \ln\!\left(  \pi\lambda\,r_{0}^{2}+1\right)
-\ln\!\left(  \pi\lambda\,r_{0}^{2}\right)  \right)  \right]  }.
\end{equation}
Note that for $\lambda\rightarrow\infty$, $\omega_{r}\left(  r_{0}\right)
\rightarrow-\infty$, as it should be. This can be easily understood by looking
at the behavior of the energy density $\rho_{4}\left(  r\right)  $ in the same
limit. On the other hand when $r\rightarrow\infty$, one finds%
\begin{equation}
\omega_{r}\left(  r\right)  \underset{r\rightarrow\infty}{\rightarrow}%
\frac{2\sqrt{\pi\lambda}}{3}\,\left(  -\pi^{\frac{3}{2}}\ln\!\left(
1+\frac{1}{r_{0}^{2}\lambda\pi}\right)  \,\lambda^{\frac{3}{2}}r_{0}^{3}%
+r_{0}\sqrt{\pi\lambda}-\pi+2\arctan\!\left(  r_{0}\sqrt{\pi\lambda}\right)
\right)  r-\frac{16r_{0}\pi\lambda}{3C^{2}\kappa}r+O\!\left(  1\right)
,\label{or(r)Mix}%
\end{equation}
namely $\omega_{r}\left(  r\right)  $ is linearly divergent for $r\rightarrow
\infty$. Like in the previous sections, we are going to modify the
construction of $\omega_{r}\left(  r\right)  $ in such a way that%
\begin{equation}
\omega_{r}\left(  r\right)  =-\frac{r_{0}b\left(  r\right)  }{r^{2}b^{\prime
}\left(  r\right)  }.
\end{equation}
We know that $\omega_{r}\left(  r_{0}\right)  $ does not change, while for
$r\rightarrow\infty$, we get%
\begin{equation}
\omega_{r}\left(  r\right)  \underset{r\rightarrow\infty}{\rightarrow}%
\frac{2r_{0}\sqrt{\pi\lambda}}{3}\,\left(  -\pi^{\frac{3}{2}}\ln\!\left(
1+\frac{1}{r_{0}^{2}\lambda\pi}\right)  \,\lambda^{\frac{3}{2}}r_{0}^{3}%
+r_{0}\sqrt{\pi\lambda}-\pi+2\arctan\!\left(  r_{0}\sqrt{\pi\lambda}\right)
\right)  -\frac{16r_{0}^{2}\pi\lambda}{3C^{2}\kappa}.\label{or(r)MMix}%
\end{equation}
Even for the mixed case, the redshift function obeys the differential equation
$\left(  \ref{Phi'(r)Ma1}\right)  $ whose solution close to the throat is
represented by Eq.$\left(  \ref{Phi(r)MT}\right)  $. Only the value of
$B=b^{\prime}(r_{0})$ represented by Eq.$\left(  \ref{b'(r0)Mix}\right)  $ is
different as it should be. On the other hand, when $r\rightarrow\infty$, we
can use the asymptotic expression of $b\left(  r\right)  $ leading to the same
analytic form of $\left(  \ref{Phi(r)MA}\right)  $, but with $b_{M,\infty}$
replaced by $b_{Mix,\infty}$. Even with this profile, we have to check if
$\Phi(r)$ satisifies the SET equation and the third EFE. It is easy to check
that also for the mixed profile $\omega_{t}\!\left(  r\right)  $ assumes the
same analytic expression of Eq.$\left(  \ref{ot(r)Ma1}\right)  $. This means
that on the throat we will obtain the same value described in Eq.$\left(
\ref{ot(r0)M}\right)  $. Only the value of $B$ will be different as it should
be. On the other hand, when $r\rightarrow\infty$, one gets%
\begin{equation}
\omega_{t}\left(  r\right)  \simeq\frac{C\kappa\pi\sqrt{\pi\lambda}%
r_{0}-2C\kappa\sqrt{\pi\lambda}r_{0}\arctan\!\left(  \sqrt{\pi\lambda}%
r_{0}\right)  +\kappa C\ln\!\left(  \pi\lambda r_{0}^{2}+1\right)  +\lambda
r_{0}^{2}}{C\kappa\ln\!\left(  \pi\lambda r^{2}\right)  }.\label{ot(r)MixL}%
\end{equation}
The same results hold also for the third EFE as it should be. Also for the
mixed energy density profile, we can observe that, although the original
$\omega_{r}\left(  r\right)  $ of Eq.$\left(  \ref{or(r)Mix}\right)  $ is
divergent when $r\rightarrow\infty$, the radial pressure given by Eq.$\left(
\ref{EoS1}\right)  $ is not. The reason is the same of the previous two
sections: the behavior of the energy density when $r\rightarrow\infty$ is
$\rho(r)\sim1/r^{2}$ leading to a radial pressure that decreases like $1/r$.
The modification $\left(  \ref{or(r)MMix}\right)  $ allows not only to have a
convergent $\omega_{r}\left(  r\right)  $, but also a radial pressure that
goes at infinity like $1/r^{2}$. Regarding the transverse pressure, we can see
that Eq.$\left(  \ref{EoS2}\right)  $ and the approximation $\left(
\ref{ot(r)MixL}\right)  $ tell us that $p_{t}\left(  r\right)  \simeq1/\left(
\ln\!\left(  r\right)  r^{2}\right)  $ when $r\rightarrow\infty$.

\section{Conclusion}
In this paper, we have considered different energy density profiles inspired
by holographic dark energies as possible sources needed to have traversable
wormhole solutions. Since in each profile the energy density is positive and
since it is the NEC that must be violated, we are forced to introduce an EoS
of the form $\left(  \ref{EoS1}\right)  $. This implies that%
\begin{equation}
\rho\left(  r\right)  +p_{r}\left(  r\right)  =\left(  1+\omega_{r}\left(
r\right)  \right)  \rho(r)\leq0
\end{equation}
which implies $\omega_{r}\left(  r\right)  <-1$. This means that our energy
density profiles are of the \textquotedblleft\textit{phantom}%
\textquotedblright\ type. With such an EoS, we have tried to impose ZTF, that
it means that $\Phi(r)$ can assume a constant value or it can be vanishing.
Unfortunately, as a size effect the function $\omega_{r}\left(  r\right)
\rightarrow\infty$ when $r\rightarrow\infty$. To overcome this problem, we
have modified the $\omega_{r}\left(  r\right)  $ function in such a way that
the behavior at infinity is convergent, while on the throat is unchanged and
well defined. With this modification every profile admits a solution
describing a TW. However, only one energy density proposal needs no
modification. This is represented by the Bekenstein-Hawking energy density
proposal which has a regular behavior at infinity and on the throat since the
beginning. Moreover, such a profile leads to a shape function which is in
agreement with that one proposed in Ref.\cite{CLR}. It is interesting also to
note that every energy density proposal is in the \textit{phantom} energy regime. Moreover, along with the present study, one can consider an extension of the HDE wormhole solutions by introducing a Yukawa deformation, see e.g., \cite{Garattini:2021kca,deOliveira:2022hew}. This way allows us to have the possibility of building a new family of solutions.

\section*{Acknowledgments}

P. Channuie is partially supported by the Thailand National Science, Research and Innovation Fund (TSRF) via PMU-B with grant No.B37G660013.


\end{document}